\documentclass[%
 reprint,
 amsmath,amssymb,
pra,
]{revtex4-1}

\usepackage{graphicx}% Include figure files
\usepackage{dcolumn}% Align table columns on decimal point
\usepackage{bm}% bold math
\newcommand{\be}{\begin{equation}}
\newcommand{\ee}{\end{equation}}
\newcommand{\ben}{\begin{eqnarray}}
\newcommand{\een}{\end{eqnarray}}

\def\ba{\begin{eqnarray}}
\def\ea{\end{eqnarray}}

\def\lb{\label}
\def\be{\begin{equation}}
\def\ee{\end{equation}}
\begin{document}
\title{2D Zak Phase Landscape in Photonic Discrete-Time Quantum Walks}

\author{Graciana Puentes$^{1,2}$}
    \email[Correspondence email address: ]{gpuentes@df.uba.ar}% Your name
    \affiliation{1-Departamento de Fsica, Facultad de Ciencias Exactas y Naturales, Universidad de Buenos Aires, Ciudad Universitaria, 1428 Buenos Aires, Argentina,\\
    2-CONICET-Universidad de Buenos Aires, Instituto de Fsica de Buenos Aires (IFIBA), Ciudad Universitaria, 1428
Buenos Aires, Argentina. 
}

\date{\today} % Leave empty to omit a date

\begin{abstract}
We present a study of the  2D  Zak phase landscape in photonic discrete-time quantum walk (DTQW) protocols. In particular, we report numerical results for three different DTQW scenarios which preserve spatial inversion symmetry (SIS) and time-reversal symmetry (TRS), while presenting a non-trivial Zak phase structure, as a consequence of a non-vanishing Berry connection. Additionally, we propose a novel approach to break TRS in photonic systems, while preserving a vanishing Berry curvature. Our results bear a close analogy to the Aharonov-Bohm effect, stating that in a field-free multiply connected region of space, the evolution of the system depends on vector potentials, due to the fact that the underlying canonical formalism cannot be expressed in terms of fields alone. 
\end{abstract}


\maketitle

\section{Introduction}

It is well known that if a quantum particle, in a given eigenstate with energy $E$, is slowly transported around a circuit $\gamma$  by varying parameters \emph{$\textbf{R}$} in its Hamiltonian \emph{$H(\textbf{R})$}, the system will acquire a geometric phase factor \emph{$\gamma (\textbf{R})$}, in addition to the standard dynamical phase factor $e^{iE t}$ \cite{Berry}. For quantum particles returning adiabatically to their initial state, while storing information about the circuit on the geometric phase, such geometric phase factor can be defined as \cite{Berry}:
\begin{equation}
e^{i\gamma}=\langle \psi_{\mathrm{ini}}|\psi_{\mathrm{final}} \rangle.
 \end{equation}
 Geometric phases can be held responsible for a number of situations: they affect material properties in solids, such as conductivity in graphene \cite{Berrygraphene}, they trigger the emergence of surface edge-states in topological insulators, whose surface electrons experience a geometric phase \cite{Berrytopoinsul}, they can modify the outcome of molecular chemical reactions  \cite{Berrychemestry}, and could even have implications for quantum information technology, via the Majorana particle \cite{Berrymayorana}, in addition to  bearing close analogies to gauge-field theories \cite{BerryGauge}. Here we present a simple system based on Discrete-Time Quantum Walk (DTQW) architectures in 2D, which presents a non-trivial  2D Berry phase on the torus, i.e., the Zak phase \cite{Zak}. Non-trivial topology is usually regarded as a consequence of a non-vanishing Berry curvature ($F$), leading to invariant Chern numbers. Nevertheless, a non-trivial topological scenario can also arise as a direct consequence of a non-vanishing Berry connection ($A$), even in the absence of Berry curvature \cite{Liu2017}. This is the type of scenario we consider in this article.\\ 

~Discrete-Time Quantum Walks (DTQWs) \cite{Aharonov} offer a versatile platform for the exploration of a wide range of non-trivial geometric and topological phenomena (experiment) \cite{Kitagawa} \cite{Crespi} \cite{Alberti}, and (theory) \cite{Kitagawa2,Obuse,Shikano2,Wojcik,MoulierasJPB}. Further, QWs are robust platforms for modelling a variety of dynamical processes from excitation transfer in spin chains \cite{Bose,Christandl} to energy transport in biological complexes \cite{Plenio}. They enable to study multi-path quantum interference phenomena \cite{bosonsampling1,bosonsampling2,bosonsampling3,bosonsampling4}, and can provide for a route to validation of quantum complexity \cite{validation1,validation2}, and universal quantum computing \cite{Childs}. Moreover, multi-particle QWs warrant a powerful tool for encoding information in an exponentially larger space, and for quantum simulations in biological, chemical and physical systems, in 1D and 2D geometries \cite{Peruzzo} \cite{SchreiberScience}. \\

In this paper, we report a simple theoretical scheme for generation  of a non-trivial geometric Zak  phase landscape in 2D DTQW architectures, based on a non-zero Berry connection. The systems preserves both spatial inversion symmetry (SIS) and time reversal symmetry (TRS), and therefore has a zero Berry curvature. Moreover, we propose a method to break TRS based on photonic quantum walks. Our results bear a close analogy to the Aharonov-Bohm effect, which essentially confirms that in a field-free ($F = 0$) multiply-connected region of space, the physical properties of the system depend on vector potentials ($A$), due to the fact that the Schroedringer equation is obtained from a canonical formalism, which cannot be expressed in terms of fields alone \cite{AharonovBohm1959}.    \\

\begin{figure}[t!]
\includegraphics[width=0.9\linewidth]{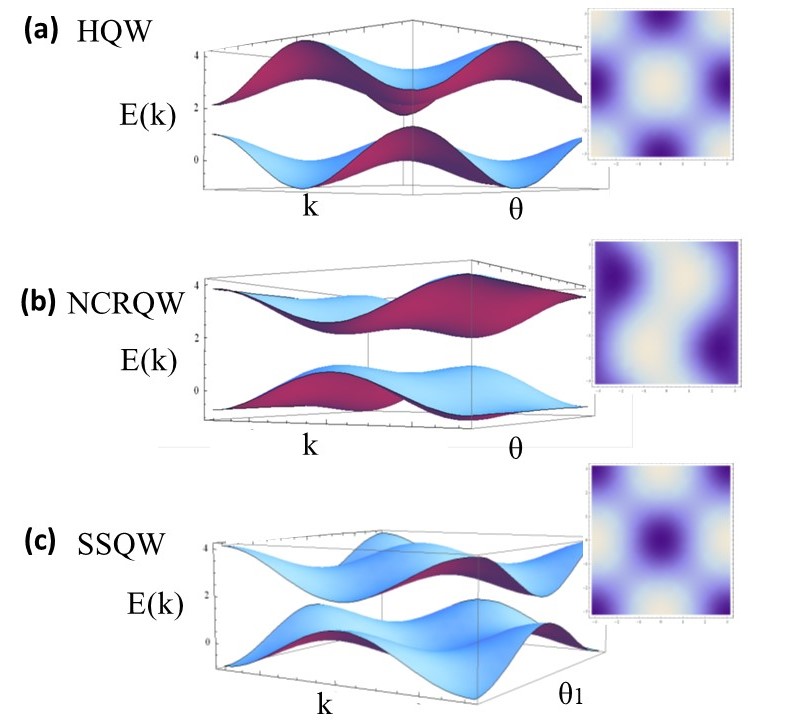} \caption{3D Energy dispersion relations as a function of quasi-momentum ($k$) and angular parameter ($\theta$) for (a) Hadamard Quantum Walk (HQW), (b) Non-Commuting Rotations Quantum Walk (NCRQW) for parameter $\phi=0$, and (c) Split Step Quantum  Walk (SSQW) for parameter $\theta_2=0$. Insets depict density plots of 3D dispersion relations. }
\end{figure}

The 2D topological landscape can be characterized  by the extended Zak phase in 2D, which is also the wave polarization vector ($\textbf{Z}$), given by \cite{Liu2017}:

\begin{equation}
\textbf{Z}=\frac{1}{2\pi} \int _{BZ} dk_{x}dk_{y} \mathrm{Tr}[\textbf{A}(k_{x},k_{y})],
\end{equation}

where $\textbf{A}= \langle u | i \partial_{\textbf{k}}| u \rangle$ is the Berry connection, and $| u \rangle$ are the Bloch eigenvectors in the first Brillouin zone (BZ), obtained by diagonalization of the Hamiltonian. Integration is performed over the first BZ. Spatial inversion symmetry (SIS) and time reversal symmetry (TRS) place strong limitations 
on the actual values that the 2D Zak phase $\textbf{Z}=(Z_{x},Z_{y})$  can take. In fact, systems with SIS and TRS are forced to have zero Berry curvature ($F=0$), where the Berry curvature is defined as the curl of the Berry connection ($F=\nabla \times \textbf{A}$), due to the fact that TRS requires $F(-\textbf{k})=-F(\textbf{k})$, while SIS requires $F(-\textbf{k})=F(\textbf{k})$, which yields $F=-F=0$.
This constraint forces ($Z_{x}=Z_{y}$). Nevertheless, in the particular photonic implementation we consider here, it is possible to break TRS, obtaining ($Z_{x}=-Z_{y}$), while the Berry curvature remains equal to zero because the Hamiltonian is separable and Bloch eigenvectors do not depend on orthogonal wavevectors, meaning $ \partial_{k_{x}}| u_{y} \rangle=\partial_{k_{y}}| u_{x} \rangle=0$ (Eq. 3.11).  \\

\section{Discrete-Time Quantum Walks (DTQW)}

The basic step in the standard 1D DTQW is given by a unitary evolution operator $U(\theta)=TR_{\vec{n}}(\theta)$, where $R_{\vec{n}}(\theta)$ is a rotation along an arbitrary direction $\vec{n}=(n_{1},n_{1},n_{3})$, given by $$R_{\vec{n}}(\theta)=
\left( {\begin{array}{cc}
 \cos(\theta)-in_{3}\sin(\theta) & (in_{1}-n_{2})\sin(\theta)  \\
 (in_{1}+n_{2})\sin(\theta) & \cos(\theta) +in_{3}\sin(\theta)  \\
 \end{array} } \right), $$in the Pauli basis \cite{Pauli}. In this basis, the y-rotation is defined by a coin operator of the form  \cite{Pauli}.
$$R_{y}(\theta)=
\left( {\begin{array}{cc}
 \cos(\theta) & -\sin(\theta)  \\
 \sin(\theta) & \cos(\theta)  \\
 \end{array} } \right). $$  This is  
followed by a spin- or polarization-dependent translation $T_{x(y)}$ given by 
$$
T_{x(y)}=\sum_{x(y)}|x(y)+1\rangle\langle x(y) | \otimes|H\rangle \langle H| +|x(y)-1\rangle \langle x(y)| \otimes |V\rangle \langle V|,
$$
 where $H=(1,0)^{T}$ and $V=(0,1)^{T}$.
The evolution operator for a discrete-time step is equivalent to that generated by a Hamiltonian $H(\theta)$, such that $U(\theta)=e^{-iH(\theta)}$ ($\hbar=1$), with $$H(\theta)=\int_{-\pi}^{\pi} dk[E_{\theta}(k)\vec{n}(k).\vec{\sigma}] \otimes |k \rangle \langle k|,$$ and $\vec{\sigma}$ the Pauli matrices, which readily reveals the spin-orbit coupling mechanism in the system.~The quantum walk described by $U(\theta)$ has been realized experimentally in a number of systems \cite{photons,photons2,photons3,ions,coldatoms}, and has been shown to posses chiral symmetry, and display Dirac-like dispersion relation given by $\cos(E_{\theta}(k))=\cos(k)\cos(\theta)$. Extension to the 2D case can be easily performed by considering independent evolution operators ($U_{x(y)}(\theta_{x(y)}$)), for each dimension ($x,y$).

 Due to spatial periodicity of the Hamiltonian, the eigenstates of the system obey the Bloch theorem and can be written as:  
\begin{equation}
|\psi_{k}(r) \rangle= e^{ik.r}|u_{k}(r) \rangle,
\end{equation}
where $|u_{k}(r) \rangle$ are the Bloch eigenvectors which obey the same periodicity of the Hamiltonian, and satisfy the eigenvalue equation:
\begin{equation}
H_{k}|u_{k}(r) \rangle=E(k)|u_{k}(r) \rangle,
\end{equation}
where $E(k)$ is the energy dispersion relation. The physics of the system is captured by the dispersion relation, and by the geometrical properties of the Bloch eigenvectors \cite{Ozawa2018}, as described in the following Sections. 

In the following, we present and characterize three different DTQW scenarios for calculation of the non-trivial geometrical Zak phase landscape in 2D. We consider a generic wavevector $k$, which eventually will indicate each of the two dimensions under consideration $(x,y)$.

\begin{figure} [t!]
\includegraphics[width=1\linewidth]{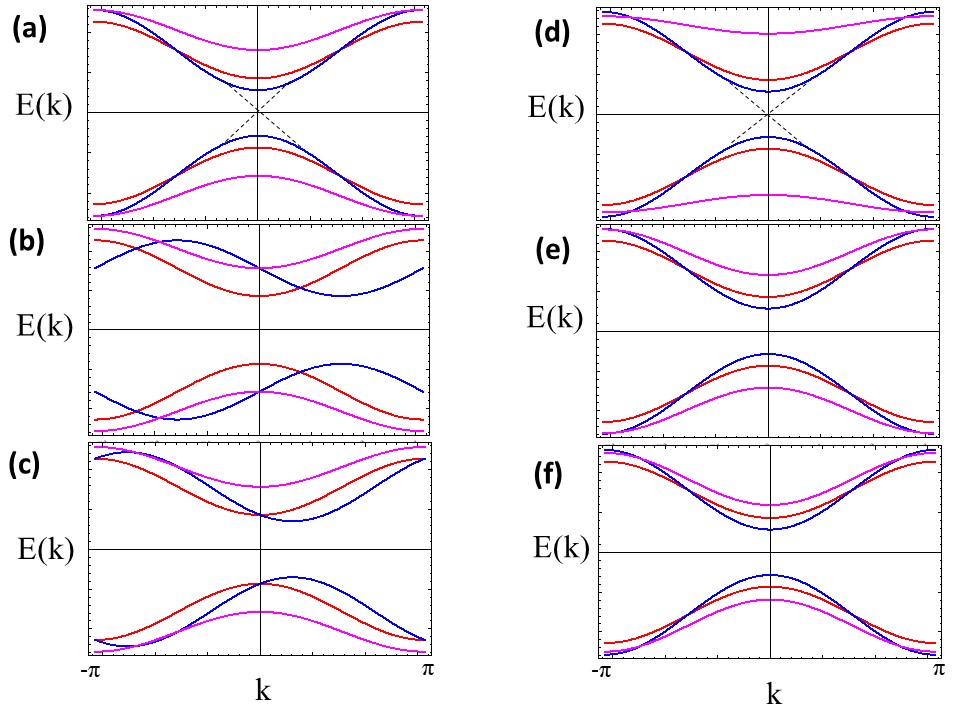} \caption{2D Energy dispersion relations as a function of quasi-momentum ($k$), for different values of the system parameters. HQW, NCRQW and SSQW are represented by red, blue and magenta curves, respectively. Dispersion relation for (a) HQW with $\theta=\pi/4$, NCRQW with ($\phi=\pi/4,\theta=0$), SSQW with ($\theta_1=\pi/4,\theta_2=\pi/4$), (b)  HQW with $\theta=\pi/4$, NCRQW with ($\phi=\pi/4,\theta=\pi/2$), SSQW with ($\theta_1=\theta_2=\Pi/4$), (c)  HQW with $\theta=\pi/4$, NCRQW with ($\phi=\theta=\pi/4$), SSQW with ($\theta_1=\theta_2=\pi/4$), (d)  HQW with $\theta=\pi/4$, NCRQW with $(\phi=\pi/4,\theta=0$), SSQW with ($\theta_1=\pi/4,\theta_2=2 \pi/5$), (e)  HQW with $\theta=\pi/4$, NCRQW with ($\phi=\pi/4,\theta=0$), SSQW with ($\theta_1=\pi/4,\theta_2=\pi/5$), and (f)  HQW with $\theta=\pi/4$, NCRQW with ($\phi=\pi/4,\theta=0$), SSQW with ($\theta_1=\pi/4,\theta_2=\pi/8$). Linear dispersions at gapless Dirac points are apparent. }
\end{figure}

\subsection{Hadamard Quantum Walk (HQW)}

The first example is the traditional Hadamard Quantum Walk \cite{Aharonov}, consisting of a rotation along the y-direction ($R_{y}(\theta)$) by an angle ($\theta$) followed by a spin-dependent translation ($T$). The 3D axes assigned to norms and rotations will be labelled by the indices ($i=1,2,3$), in  order to distinguish them from the 2D (spatial or temporal) dimensions $(x,y)$ for implementation of the DTQW. The 3D-norm for decomposing the quantum walk Hamiltonian of the system in terms of Pauli matrices $H_{\mathrm{QW}}=E(k)\vec{n} \cdot \vec{\sigma}$  becomes \cite{Kitagawa}: \\

\begin{equation}
\begin{array}{ccc}
n_{\theta}^{1}(k)&=&\frac{\sin(k)\sin(\theta)}{\sin(E_{\theta}(k))}\\
n_{\theta}^{2}(k)&=&\frac{\cos(k)\sin(\theta)}{\sin(E_{\theta}(k))}\\
n_{\theta}^{3}(k)&=&\frac{-\sin(k)\cos(\theta)}{\sin(E_{\theta}(k))},\\
\end{array}
 \end{equation}
where $k$ represents the wavevector in either dimension $(x,y)$. The dispersion relation for the Hadamard quantum walk results in \cite{Kitagawa2}:
$$
 \cos(E_{\theta}(k))=\cos(k)\cos(\theta),
$$
which corresponds to a Dirac-like, i.e., a linear dispersion relation when the gap is closed at Dirac points \cite{Kitagawa2}. 3D and 2D Plots of the dispersion relation characterizing the DTQWs are displayed in Figure 1 and Figure 2.

\subsection{Non-Commuting Rotations Quantum Walk (NCRQW)}

The second example consists of a DTQW based on two consecutive non-commuting rotations followed by a spin-dependent translation \cite{PuentesCrystal,PuentesEntropy}. The first rotation ($R_{2}(\theta)$)  is performed along the y-direction ($i=2$) by an angle ($\theta$), and the second rotation ($R_{1}(\phi)$) is performed along the x-direction ($i=1$) by an angle $\phi$, such that the unitarity step becomes $U(\theta,\phi)=TR_{1}(\phi)R_{2}(\theta)$, where $R_{1}(\phi)$ is given in the same basis \cite{Pauli} by:
$$R_{1}(\phi)=
\left( {\begin{array}{cc}
 \cos(\phi) & i\sin(\phi)  \\
i \sin(\phi) & \cos(\phi)  \\
 \end{array} } \right).$$ The modified dispersion relation becomes:

$$
\cos(E_{\theta,\phi}(k))= \cos(k)\cos(\theta)\cos(\phi) +\sin(k)\sin(\theta)\sin(\phi), 
$$

 where we recover the Hadamard dispersion relation for $\phi=0$, as expected.
~The 3D-norm for decomposing the Hamiltonian of the system in terms of Pauli matrices  becomes: 
\begin{equation}
\begin{array}{ccc}
n_{\theta,\phi}^{1}(k)&=&\frac{-\cos(k)\sin(\phi)\cos(\theta)+\sin(k)\sin(\theta)\cos(\phi)}{\sin(E_{\theta,\phi}(k))}\\
n_{\theta,\phi}^{2}(k)&=&\frac{\cos(k)\sin(\theta)\cos(\phi)+\sin(k)\sin(\phi)\cos(\theta)}{\sin(E_{\theta,\phi}(k))}\\
n_{\theta,\phi}^{3}(k)&=&\frac{-\sin(k)\cos(\theta)\cos(\phi)+\cos(k)\sin(\theta)\sin(\phi)}{\sin(E_{\theta,\phi}(k))}.\\
\end{array}
 \end{equation}

Dispersion relations for the DTQW with two consecutive non-commuting rotations within the first Brillouin zone are of the Dirac type (linear) at the set of gapless Dirac points, where the quasi-energy gap closes at $E(k)=0$, as displayed in Figure 1 and Figure 2. The second rotation enables to close the gap at zero energy for complementary points, and allows to create a non-trivial geometric Zak phase structure in the system. In particular, this system has a non-trivial phase diagram with a larger number of gapless points for different momenta as compared to the system consisting of a single rotation. We calculated analytically the gapless Dirac points for the system. Using basic trigonometric considerations, it can be shown that the energy gap closes at 13 discrete points, for different values of quasi-momentum $k$ \cite{PuentesCrystal,PuentesJOSAB,PuentesEntropy}. \\

\subsection{Split-Step Quantum Walk (SSQW)}

The third DTQW protocol consists of two consecutive spin-dependent translations $T$ and rotations $R$, such that the unitary step becomes $U(\theta_1,\theta_2)=TR(\theta_1)TR(\theta_2)$, as described in detail in \cite{Kitagawa2}. The so-called ``Split-Step" Quantum Walk (SSQW), has been shown to possess a non-trivial topological landscape characterized by topological sectors with different topological numbers, such as the winding number $W=0,1$. The dispersion relation for the split-step quantum walk results in \cite{Kitagawa2}:

%\begin{equation}
%\begin{array}{ccc}
% \cos(E_{\theta_1,\theta_2}(k))&=& \\                     
%&=& \cos(k)\cos(\theta_1)\cos(\theta_2)-\sin(\theta_1)\sin(\theta_2).\\%
%\end{array}
%\end{equation}

$$
\cos(E_{\theta_1,\theta_2}(k))=\cos(k)\cos(\theta_1)\cos(\theta_2)-\sin(\theta_1)\sin(\theta_2).
$$

The 3D-norm for decomposing the quantum walk Hamiltonian of the system in terms of Pauli matrices $H_{\mathrm{QW}}=E(k)\vec{n} \cdot \vec{\sigma}$  becomes \cite{Kitagawa}: \\
\begin{equation}
\begin{array}{ccc}
n_{\theta_1,\theta_2}^{1}(k)&=&\frac{\sin(k)\sin(\theta_1)\cos(\theta_2)}{\sin(E_{\theta_1,\theta_2}(k))}\\
n_{\theta_1,\theta_2}^{2}(k)&=&\frac{\cos(k)\sin(\theta_1)\cos(\theta_2)+\sin(\theta_2)\cos(\theta_1)}{\sin(E_{\theta_1,\theta_2}(k))}\\
n_{\theta_1,\theta_2}^{3}(k)&=&\frac{-\sin(k)\cos(\theta_2)\cos(\theta_1)}{\sin(E_{\theta_1,\theta_2}(k))}.\\
\end{array}
 \end{equation}
The dispersion relation and topological landscape for the split-step quantum walk was analysed in detail in \cite{Kitagawa2}.  \\

Figure 1 displays 3D energy dispersion relations as a function of the quasi-momentum ($k$) and the angular parameter ($\theta$) within the first Brillouin zone, for (a) Hadamard Quantum Walk (HQW), (b) Non-Commuting Rotations Quantum Walk (NCRQW) considering ($\phi=0$), and (c) Split Step Quantum  Walk (SSQW) considering ($\theta_2=0$).\\

Figure 2 displays 2D Energy dispersion relations as a function of the quasi-momentum ($k$), for different values of the system parameters. HQW, NCRQW and SSQW are represented by red, blue and magenta curves, respectively. Dispersion relation for (a) HQW with $\theta=\pi/4$, NCRQW with ($\phi=\pi/4,\theta=0$), SSQW with ($\theta_1=\pi/4,\theta_2=\pi/4$), (b)  HQW with $\theta=\pi/4$, NCRQW with ($\phi=\pi/4,\theta=\pi/2$), SSQW with ($\theta_1=\theta_2=\Pi/4$), (c)  HQW with $\theta=\pi/4$, NCRQW with ($\phi=\theta=\pi/4$), SSQW with ($\theta_1=\theta_2=\pi/4$), (d)  HQW with $\theta=\pi/4$, NCRQW with $(\phi=\pi/4,\theta=0$), SSQW with ($\theta_1=\pi/4,\theta_2=2 \pi/5$), (e)  HQW with $\theta=\pi/4$, NCRQW with ($\phi=\pi/4,\theta=0$), SSQW with ($\theta_1=\pi/4,\theta_2=\pi/5$), and (f)  HQW with $\theta=\pi/4$, NCRQW with ($\phi=\pi/4,\theta=0$), SSQW with ($\theta_1=\pi/4,\theta_2=\pi/8$). Linear dispersions at gapless Dirac points are apparent.

\begin{figure} 
\includegraphics[width=0.9\linewidth]{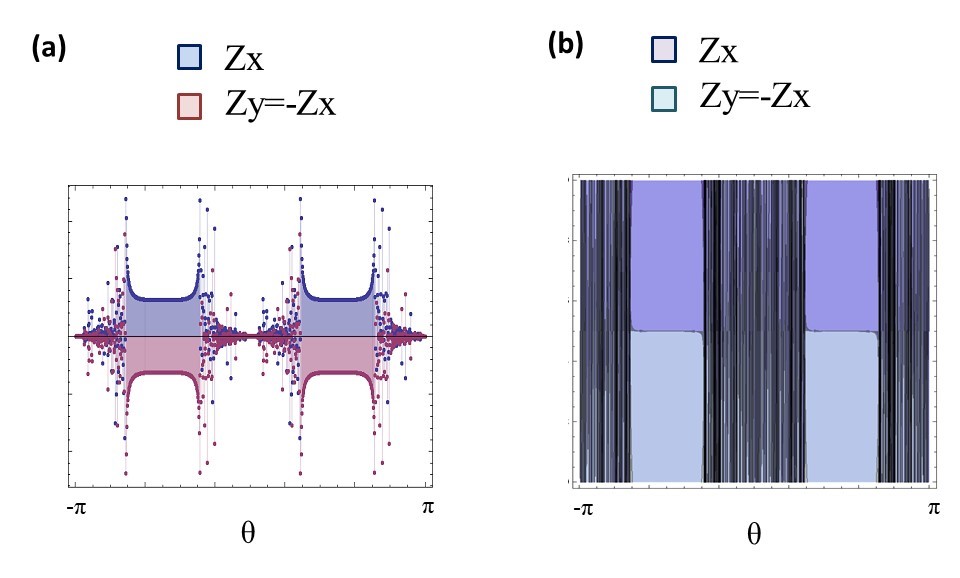} \caption{2D Zak phase landscape ($Z_{x}, Z_{y}$) for Hadamard Quantum Walk. (a) Blue curve corresponds to Zak phase in x-direction ($Z_{x}$) and purple curve corresponds to Zak phase in y-direction ($Z_{y}$). By switching the Bloch vector arguments  $\phi(k_{x})=-\phi(k_{y})$ characterizing the quantum walk in x-direction and quantum walk in y-direction it is possible to break time reversal symmetry ($TRS=-1$), obtaining  ($Z_{x}=-Z_{y}$). (b) Contour plot of $Z_{x}$ and $Z_{y}$.  }
\end{figure}

\section{Zak Phase Calculation}

We will now give expressions for the 2D Zak Phase in the different scenarios presented in the previous Section. These scenarios are casted by the following general Hamiltonian, either in the x-dimension or the y-dimension which, in turn, is specified by the wave-number ($k_{x}, k_{y}$):

\begin{equation}
H\sim n_1 \sigma_1+n_2 \sigma_2+ n_3 \sigma_3.
\end{equation}

The Hamiltonians to be described differ by a multiplying scalar  factor corresponding to the quasi-energy ($E(k)$), and by the actual expression of the 3D-norm $n_i$ ($i=1,2,3$). But since the Bloch eigenvectors are the only quantities of interest for the present problem, the overall constants of this Hamiltonian can be safely ignored. Now, our generic hamiltonian is given by the matrix
 \be
H=
\left(
\begin{array}{cc}
  n_3 \qquad    n_1-in_2\\
 n_1+i n_y \qquad   -n_3   
\end{array}
\right),
\ee
and has the following eigenvalues
\begin{equation}
\lambda=\pm \sqrt{n_1^2+n_2^2+n_3^2}
\end{equation}
The normalized eigenvectors, also called the Bloch eigenvector,  in either x-direction or y-direction, then result
\begin{equation}
|u_\pm>= 
\left(
\begin{array}{cc}
  \frac{n_1+i n_2}{\sqrt{2n_1^2+ 2n_2^2+2n_3^2\mp 2n_3\sqrt{n_1^2+n_2^2+n_3^2}}}    \\
  \frac{n_z\mp \sqrt{n_1^2+n_2^2+n_3^2}}{\sqrt{2n_1^2+ 2n_2^2+2n_3^2\mp 2n_z\sqrt{n_1^2+n_2^2+n_3^2}}} 
\end{array}
\right),
\end{equation}

where the given direction is specified by the wave-number $k_{x,y}$. This expression readily confirms that the Berry curvature ($F$) is always equal to zero, because the Bloch eigenvectors do not depend on orthogonal wavevectors, meaning $ \partial_{k_{x}}| u_{y} \rangle=\partial_{k_{y}}| u_{x} \rangle=0$, as anticipated.  Note that the scalar factor $n_i\to\lambda n_i$ does not modify the result, as expected. This results from the fact that two Hamiltonians with differ by a constant should have the same eigenvectors.

Thus, the 1D Zak phase for the problem to be considered is \cite{Zak}:
$$
Z_{x(y)}=i\int_{-\pi/2}^{\pi/2} (<u_+|\partial k_{x(y)} u_+>+ <u_-|\partial k_{x(y)} u_->) dk.
$$

As anticipated in the introduction, the 2D Zak phase can be extended in the form $\textbf{Z}=(Zx,Zy)$ (Eq. 1.2), for systems which preserve TRS and SIS. Moreover, for the particular photonic DTQW implementation we consider here, it is possible to break TRS. Namely, by switching the Bloch eigenvector arguments  $\phi(k_{x})=-\phi(k_{y})$ characterizing the quantum walk in x-direction or in y-direction, it is possible to break time-reversal symmetry ($TRS=-1$), obtaining  ($Z_{x}=-Z_{y}$), as explained in detail in Section IV. We will now apply these concepts to some specific examples.

\subsection{Zak phase for SSQW}

We first consider the Split-Step Quantum Walk. This corresponds to a quantum walk with unitary step given by $U(\theta_1, \theta_2)=TR(\theta_1)TR(\theta_2)$, as proposed in \cite{Kitagawa2}. Note that the rotations are performed around the y-direction, therefore in this case angular labels $(1,2)$ do not correspond to the cartesian axes \cite{Kitagawa2}. %In this example the 3D-norm components $n_i$ are of the following form:

%\begin{equation}
%\begin{array}{ccc}
%n_{\theta_1,\theta_2}^{1}(k)&=&\frac{\sin(k)\sin(\theta_1)\cos(\theta_2)}{\sin(E_{\theta_1,\theta_2}(k))}\\
%n_{\theta_1,\theta_2}^{2}(k)&=&\frac{\cos(k)\sin(\theta_1)\cos(\theta_2)+\sin(\theta_2)\cos(\theta_1)}{\sin(E_{\theta_1,\theta_2}(k))}\\
%n_{\theta_1,\theta_2}^{3}(k)&=&\frac{-\sin(k)\cos(\theta_2)\cos(\theta_1)}{\sin(E_{\theta_1,\theta_2}(k))}.\\
%\end{array}
% \end{equation}

We consider the particular case that $n_3=0$. By taking one of the angle parameters such that $n_3=0$, it follows that the Bloch eigenvectors of the Hamiltonian are versors, of the form \cite{PuentesJOSAB}:

\begin{equation}
|u_\pm>= 
\frac{1}{\sqrt{2}}\left(
\begin{array}{cc}
  e^{i\phi(k)}   \\
  \mp 1 
\end{array}
\right),\qquad 
\tan\phi(k)=\frac{n_2}{n_1}.
\end{equation}
There are two choices for $n_3=0$, which are $\theta_1=0$ or $\theta_2=0$.
In both cases the Zak phase results in \cite{PuentesCrystal,PuentesJOSAB}:
\begin{equation}
Z=Z_{+} +Z_{-}=i\int_{-\pi/2}^{\pi/2}dk <u_{+}|\partial_{k} u_{+}>
\end{equation}
\begin{equation}
+i\int_{-\pi/2}^{\pi/2} dk <u_-|\partial_k u_->=\phi(-\pi/2)-\phi(\pi/2),
\end{equation}
from where it follows that \cite{PuentesCrystal,PuentesJOSAB}:
\begin{equation}
Z=\frac{\tan(\theta_2)}{\tan(\theta_1)}.
\end{equation}

\begin{figure} [b!]
\includegraphics[width=1\linewidth]{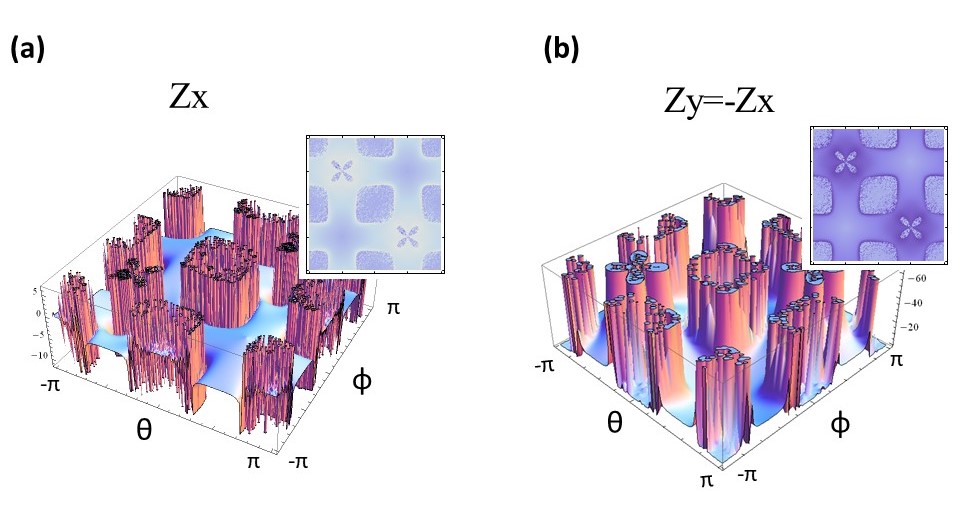} \caption{2D Zak phase landscape ($Z_{x}, Z_{y}$) for Non-Commuting Rotations Quantum Walk (NCRQW). (a) Zak phase in x-direction ($Z_{x}$) for system parameters $\theta \in [-\pi,\pi]$ and $\phi \in [-\pi,\pi]$, (b) Zak phase in y-direction ($Z_{y}$) for system parameters $\theta \in [-\pi,\pi]$ and $\phi \in [-\pi,\pi]$. By switching the Bloch vector arguments  $\phi(k_{x})=-\phi(k_{y})$ characterizing the quantum walk in x-direction and quantum walk in y-direction it is possible to break time reversal symmetry ($TRS=-1$), obtaining  ($Z_{x}=-Z_{y}$). Insets display density plots of the 3D Zak phase.  }
\end{figure}

\begin{figure}  [b!]
\includegraphics[width=1\linewidth]{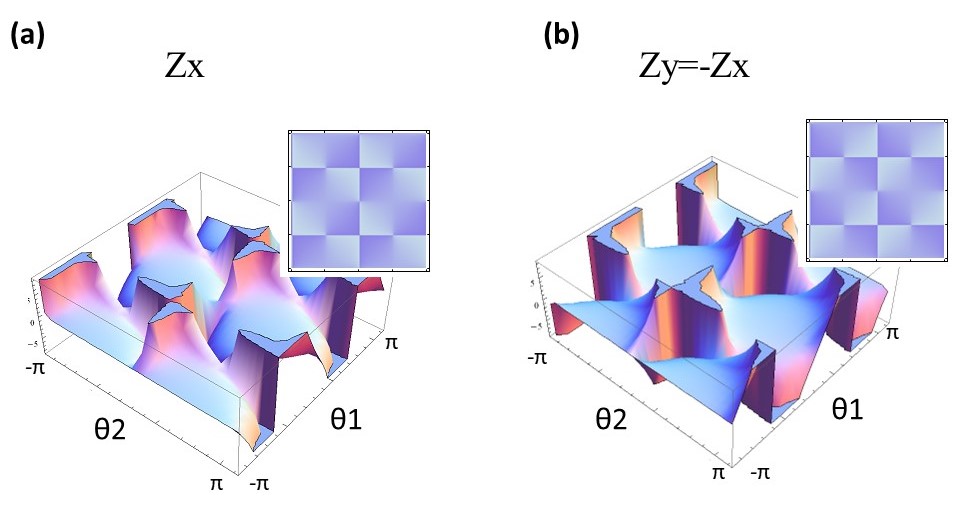} \caption{2D Zak phase landscape ($Z_{x}, Z_{y}$) for Split Step Quantum Walk (SSQW). (a) Zak phase in x-direction ($Z_{x}$) for system parameters $\theta_1 \in [-\pi,\pi]$ and $\theta_2 \in [-\pi,\pi]$, (b) Zak phase in y-direction ($Z_{y}$) for system parameters $\theta_1 \in [-\pi,\pi]$ and $\theta_2 \in [-\pi,\pi]$. By switching the Bloch vector arguments  $\phi(k_{x})=-\phi(k_{y})$ characterizing the quantum walk in x-direction and quantum walk in y-direction it is possible to break time reversal symmetry ($TRS=-1$), obtaining  ($Z_{x}=-Z_{y}$). Insets display density plots of the 3D Zak phase. }
\end{figure}

\subsection{Zak Phase for NCRQW}

Now we proceed to calculate the Zak phase for DTQW with non-commuting rotations. The unitary step as described in the introduction results in:

$$
U(\theta,\phi)=TR_{1}(\phi)R_{2}(\theta).
$$

The norms $n_i$ are of the following form
$$
n_1=-\cos(k) a+\sin(k)b,
 n_2=\cos(k) b+\sin(k) a,
$$
\be
n_3=\cos(k) c-\sin(k) d,
\ee
with
$$
a=\sin(\phi)\cos(\theta),
$$
\be\lb{ang}
 b=\cos(\phi)\sin(\theta),
 \ee
$$
 c=\sin(\phi)\sin(\theta),\qquad d=\cos(\phi)\cos(\theta).
$$
the  angular functions defined above. The numerator $C_1$ is given by

\begin{equation}
C_1=n_1+in_2=- \exp(-i k)(a-i b),
\end{equation}

in addition, $C_2$ is
$$
C_2=n_1\mp \sqrt{n_1^2+n_2^2+n_3^2}=\cos(k) c-\sin(k) d
$$
\be\lb{n2}
\mp \sqrt{a^2+b^2+c^2 \cos^2(k)+d^2 \sin^2(k)-\sin(2k)cd}.
\ee
On the other hand, the denominator  $D$ is reduced to 
$$
D_{\pm}=\sqrt{2n_1^2+ 2n_2^2+2n_3^2\mp 2n_z\sqrt{n_1^2+n_2^2+n_3^2}}
$$
$$
=\bigg(a^2+b^2+c^2 \cos^2(k)+d^2 \sin^2(k)-\sin(2k)cd
$$
$$
\mp (\cos(k) c-\sin(k) d)
$$
\be\lb{d}
\times \sqrt{a^2+b^2+c^2 \cos^2(k)+d^2 \sin^2(k)-\sin(2k)cd}\bigg)^{\frac{1}{2}}.
\ee
Considering these expressions, eigenvectors can be written  as
\begin{equation}
|u_\pm>= 
\left(
\begin{array}{cc}
 \frac{C_1}{D_\pm}    \\
  \frac{C_2}{D_\pm} 
\end{array}
\right),\qquad 
<u_\pm|=\bigg(\frac{C^\ast_1}{D_\pm},\; \frac{C_2}{D_\pm}\bigg).
\end{equation}
Therefore, calculation of the Zak phase implies
$$
Z=Z_+ +Z_-=i\int_{-\pi/2}^{\pi/2}dk <u_+|\partial_k u_+>
$$
\begin{equation}
+i\int_{-\pi/2}^{\pi/2} dk <u_-|\partial_k u_->,
\end{equation}
requiring the following quantities
$$
Z_\pm=i\int \bigg(\frac{C_1^\ast}{D_\pm^2}\partial_k C_1+\frac{C_2}{D_\pm^2}\partial_k C_2
$$
\begin{equation}
-\frac{(|C_1|^2+|C_2|^2)}{D_\pm^3}\partial_k D_\pm\bigg) dk.
\end{equation}
This expression can be simplified further \cite{PuentesCrystal},

%Due to Eq. (3.19), it follows that the first term is real.  However, an inspection of (\ref{n2}) shows that the last two terms are purely imaginary. 
%Since the overall phase should be real, it follows that these terms should cancel. This can be seen by taking into account that:
%\be\lb{fb}
%D_\pm=\sqrt{|C_1|^2+|C_2|^2},\qquad \partial_k |C_1|^2=0,
%\ee
%together with the fact that $N_2$ is real. Then  
%%\begin{equation}
%\partial_k D_\pm=\frac{2C_2\partial_k C_2}{2D_\pm},
%\end{equation}
%where (\ref{fb}) has been taken into account. Therefore
%$$
%Z_\pm=i\int \bigg(\frac{C_1^\ast}{D_\pm^2}\partial_k C_1+\frac{C_2}{D_\pm^2}\partial_k C_2
%$$
%\begin{equation}
%-\frac{(|C_1|^2+|C_2|^2)}{D_\pm^4}C_2\partial_k C_2\bigg) dk,
%\end{equation}
%but since  $D_\pm^2=|C_1|^2+|C_2|^2$ a simple calculation shows that the last two terms cancel each other. 

resulting in 
\begin{equation}
Z_\pm=i\int \frac{C_1^\ast\partial_k C_1}{D_\pm^2}dk.
\end{equation}
By taking into account  (3.18 and \ref{d}) the phases are expressed as
\begin{equation}
Z_\pm=\int \frac{|C_1|^2}{D_\pm ^2}dk=\int_0^\pi \frac{(a^2+b^2)dk}{D_\pm^2}.
\end{equation}
 
We note that in this example the case $n_{3}=0$ is completely different than in the previous case, as it returns a trivial Zak phase $Z=\pi$, since the $k$-dependence vanishes. In addition, as opposed to the SSQW where analytic expressions for the Zak phase could be elaborated, for this system the Zak phase landscape can only be obtained via numerical integration. In particular, at the gapless Dirac points the Zak phase is not defined.  Therefore, such singular points can be regarded as topological defects of dimension zero \cite{PuentesCrystal}.\\
 
\subsection{Zak phase for HQW}

Expressions for the Zak phase in the Hadamard Quantum Walk can be easily obtained by notating that the HQW is equivalent to the DTQW with non-commuting rotations for the particular case of $\phi=0$. Therefore, the final expression for the Zak Phase in the HQW results in: 

\begin{equation}
Z_\pm=\int \frac{|C_1|^2}{D_\pm ^2}dk=\int_0^\pi \frac{(a^2+b^2)dk}{D_\pm^2}, 
\end{equation}

with $C_1=n_1+in_2$, $C_2=n_1\mp \sqrt{n_1^2+n_2^2+n_3^2}$, $D_\pm=\sqrt{|C_1|^2+|C_2|^2}$,  $a=0$, $b=\sin(\theta)$, $c=0$, and $d=\cos(\theta)$.\\

%\begin{figure} 
%\includegraphics[width=0.9\linewidth]{Diapositiva1.jpg} \caption{2D Zak phase landscape for Hadamard Quantum Walk (HQW). (a) Zak phase in x-direction, %(b) Zak phase in y-direction. }
%\end{figure}

%\begin{figure} 
%\includegraphics[width=0.9\linewidth]{Diapositiva2.jpg} \caption{2D Zak phase landscape for Hadamard Quantum Walk (HQW). (a) Zak phase in x-direction, %(b) Zak phase in y-direction. }
%\end{figure}

%\begin{figure} 
%\includegraphics[width=0.9\linewidth]{Diapositiva3.jpg} \caption{2D Zak phase landscape for Hadamard Quantum Walk (HQW). (a) Zak phase in x-direction, (b) Zak phase in y-direction. }
%\end{figure}

Figure 3 depicts 2D Zak phase landscape ($Z_{x}, Z_{y}$) for Hadamard Quantum Walk. Figure 3 (a) Blue curve corresponds to Zak phase in x-direction ($Z_{x}$) and purple curve corresponds to Zak phase in y-direction ($Z_{y}$). By switching the Bloch vector arguments  $\phi(k_{x})=-\phi(k_{y})$ characterizing the quantum walk in x-direction and quantum walk in y-direction it is possible to break time reversal symmetry ($TRS=-1$), obtaining  ($Z_{x}=-Z_{y}$). Figure 3 (b) Contour plot of $Z_{x}$ and $Z_{y}$.  \\

Figure 4 depicts 2D Zak phase landscape ($Z_{x}, Z_{y}$) for Non-Commuting Rotations Quantum Walk (NCRQW). (a) Zak phase in x-direction ($Z_{x}$) for system parameters $\theta \in [-\pi,\pi]$ and $\phi \in [-\pi,\pi]$, (b) Zak phase in y-direction ($Z_{y}$) for system parameters $\theta \in [-\pi,\pi]$ and $\phi \in [-\pi,\pi]$. By switching the Bloch eigenvector arguments  $\phi(k_{x})=-\phi(k_{y})$ characterizing the quantum walk in x-direction and quantum walk in y-direction, it is possible to break time reversal symmetry ($TRS=-1$), obtaining  ($Z_{x}=-Z_{y}$). Insets display density plots of the 3D Zak phase.\\

Figure 5 depicts 2D Zak phase landscape ($Z_{x}, Z_{y}$) for Split Step Quantum Walk (SSQW). (a) Zak phase in x-direction ($Z_{x}$) for system parameters $\theta_1 \in [-\pi,\pi]$ and $\theta_2 \in [-\pi,\pi]$, (b) Zak phase in y-direction ($Z_{y}$) for system parameters $\theta_1 \in [-\pi,\pi]$ and $\theta_2 \in [-\pi,\pi]$. By switching the Bloch eigenvector arguments  $\phi(k_{x})=-\phi(k_{y})$ characterizing the quantum walk in x-direction and quantum walk in y-direction, it is possible to break time reversal symmetry ($TRS=-1$), obtaining  ($Z_{x}=-Z_{y}$). Insets display density plots of the 3D Zak phase.\\

\section{Time-Reversal Symmetry (TRS)}

Now we propose a simple method to break time-reversal symmetry (TRS), enabling a 2D Zak phase of the form ($Z_{y}=-Z_{x}$). At first, one could be tempted to change the sign of the Zak phase by imprinting a dynamical relative phase factor between the x- and y-evolutions. Nevertheless, it should be noted that a dynamical phase would not modify the acquired geometrical phase \cite{BerryPRSL1984}. In order to modify the relative sign of the Zak phase, it is required to modify the relative sign of the argument ($\phi(k_{x})=-\phi(k_{y})$) in the Bloch eigenvectors (Eq. 3.12), for each dimension ($x,y$). Let us remind that the argument of Bloch vector in each dimension  ($j=x,y$) takes  the form:

\begin{equation}
\phi(k_{j})=\arctan [\frac{n_2(k_{j})}{n_1(k_{j})}].
\end{equation}

By noting that  $\mathrm{arctan}[x]$ is an odd function of $x$, it is sufficient to modify the sign of the numerator, or the sign of the denominator, in order to change the sign of $\phi(k_{j})$.

\subsection{TRS Breaking in Split-Step Quantum Walk (SSQW)}

For the case of the SSQW, expressions for the norm components are given by (Eq. 2.7): $$n_1=\frac{\sin(k)\sin(\theta_1)\cos(\theta_2)}{\sin(E(k))},$$ and $$n_2=\frac{\cos(k)\sin(\theta_1)\cos(\theta_2)+\sin(k)\sin(\theta_1)\cos(\theta_2)}{\sin(E(k))}.$$ Note that the energy dispersion $E(k)$ cancels out of the expression for the argument $\phi(k)$.  It is straightforward to modify the sign of $n_1$, simply by inverting the sign of $\theta_1$, in the manner $\theta_1^{x}=-\theta_1^{y}$, due to the fact that $\sin(\theta_1)$ is an odd function. Note that since the sign of $\theta_1^{y}$ is inverted for \emph{all} values of $\theta_1^{x}$, this is not equivalent to a conditional operation, where the state of the coin in y-direction is modified conditioned on the state of the coin in the x-direction. Therefore, the protocol remains separable in (x,y) dimensions. Moreover, in order to ensure a relative change of sign in $\phi(k)$, $n_2$ should not change sign at the same time as $n_1$. This restricts the values of $\theta_2$ and $k$ than can be used to switch the sign of the argument $\phi(k)$, resulting in the following condition:

\begin{equation}
\frac{\tan(\theta_2)}{\tan(\theta_1)} > \cos(k).
\end{equation}

Similar conditions can be derived for all three different DTQW protocols. As an illustrative example, a plot of the allowed region of values in parameter space ($\theta_2$ and $k$), which would enable to break TRS in the SSQW, for $\theta_1=\pi/8$ and $\theta_1=\pi/4$, are indicated in Figure 6. As explained in detail in the following Section, independently switching the sign of $\phi(k)$ in either x- or y-dimension could be accomplished by using fast-switching Electro-Optic Modulators (EOMs) (Section V).

\begin{figure}
\includegraphics[width=1\linewidth]{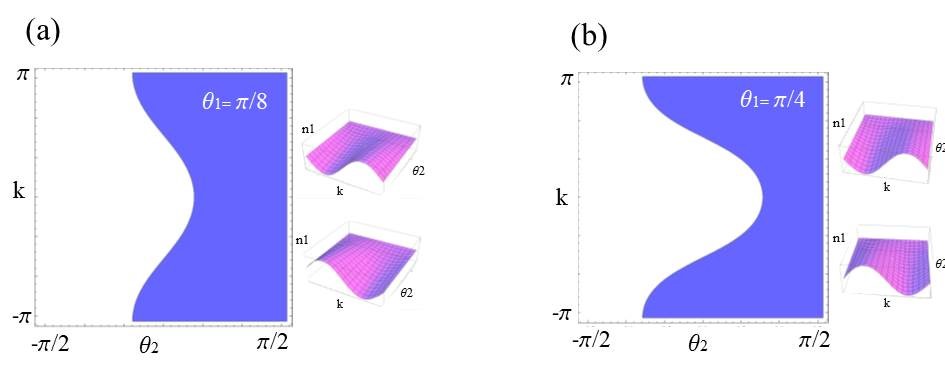} \caption{Numerical simulation displaying the allowed region (blue) of values to break time-reversal symmetry (TRS) in parameter space, for (a) $\theta_1=\pi/8$, (b) $\theta_1=\pi/4$. Insets depict 3D plots of $n_1$ confirming its sign inversion under $\theta_1 \rightarrow -\theta_1$. }
\end{figure}

\section{Proposed Experimental Scheme}

\begin{figure}[t!]
\includegraphics[width=1\linewidth]{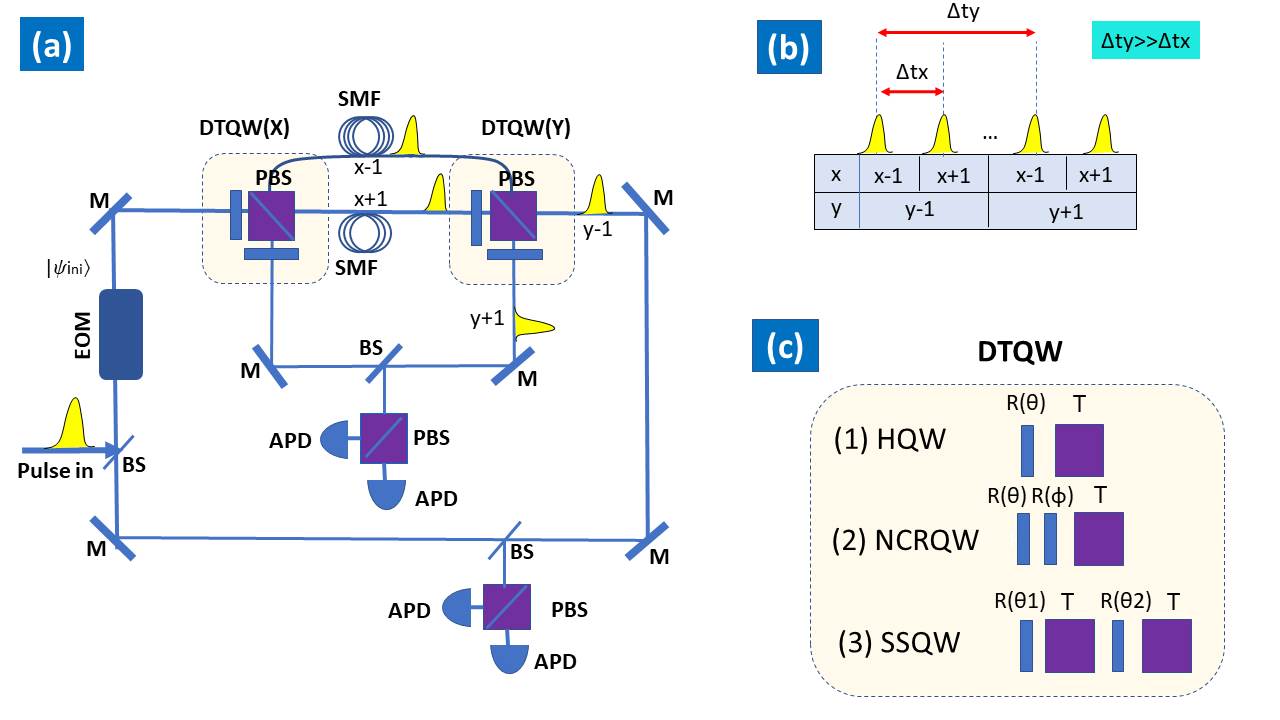} \caption{Proposed experimental scheme for implementation of 2D time-multiplexed photonic Discrete-Time Quantum  Walks  (DTQWs) based on \cite{SchreiberScience}. (a) The 2D quantum walk lattice is temporally encoded in time-multiplexed pulses with different time delays ($\Delta t_{x}$ and $\Delta t_{y}$), implemented by Single Mode Fiber (SMF) loops of adjustable lengths. Both the initial state ($|\psi_{\mathrm{ini}} \rangle$), and the quantum walk parameters $(\theta_{1},\theta_{2})$ or $(\theta,\phi)$ can be independently manipulated using fast EOMs \cite{SchreiberScience}. (b) Projection of the 2D spatial lattice onto a 1D temporally encoded pulse
chain, for step one. Time delay $\Delta t_{x}$ is adjusted by tuning the SMF length difference $|L_{2}-L_{1}|$, while $\Delta t_{y}$ is determined by the length $L_{1,2}$. Choosing $ \Delta t_{y} >> \Delta t_{x}$ it is possible to generate time bins characterized by distinctive  arrival times.  (c) Three different DTQW protocols that can be implemented using a similar experimental setup: (I) Hadamard Quantum Walk (HQW), (II) Non-Commuting Rotations Quantum Walk (NCRQW), and (III) Split-Step Quantum Walks (SSQW). Further details are in the text.  }
\end{figure}

The proposed experimental scheme is based on the novel experiment reported in Ref. \cite{SchreiberScience}, for implementation of photonic time-multiplexed DTQWs, as depicted in Fig. 7. Figure 7 (a) the 2D quantum walk lattice is temporally encoded in time-multiplexed pulses with different time delays ($\Delta t_{x}$ and $\Delta t_{y}$), implemented by Single Mode Fibres (SMF) loops of adjustable lengths. The single-photon source is an attenuated pulsed laser, typically at a wavelength of 800 nm, with a pulse width $\approx$ 90 ps, and a repetition rate of 110 kHz \cite{SchreiberScience}.
Photons are coupled into the setup through a low-reflectivity Beam Sampler (BS), and are prepared in the initial polarization state $| \psi_{\mathrm{ini}} \rangle$ using a fast Electro-Optic Modulator (EOM), corresponding to a 2D DTQW lattice initial  state $| \psi_{\mathrm{ini}} \rangle=|x, y\rangle=|0, 0 \rangle$. Upon propagation, photonic wavepackets are subject to a first step of the 1D DTQW along the x-direction. For simplicity, in Fig. 7 we consider implementation of the Hadamard Quantum Walk (HQW) in both x- and y-direction, consisting of a rotation by a HWP (coin operation) and a split by a polarizing beam splitter (PBS), although other DTQW protocols could also be implemented, as illustrated in Fig. 7 (c). In order to adjust the rotation operators
independently at each point, as required for the implementation of separable coin operations, a fast-switching Electro-Optic Modulator (EOM) can be used \cite{SchreiberScience}. 
After the first step in the x-direction, photons are routed through single-mode fibers (SMFs) of lengths ($L_1$ and $L_2$), implementing a temporal step in the y-direction. The SMF length difference  $|L_1 - L_2|$ determines the delay $\Delta t_{x}$, while the overall delay introduced by the SMFs ($L_{1,2}$) determines  $\Delta t_{y}$. Additional HWPs and a second PBS perform a
1D DTQW step in the y-direction based on the same principle. The orientation of the HWPs at each step determines the probability for the wave-packet to be translated to $x(y)-1$ or $x(y)+1$ lattice positions. Upon propagation, the photons are detected by polarization-resolving
detection of their arrival time via four avalanche photodiodes (APDs), which enables to map out the probability distribution for each lattice position (i.e., the photon statistics). Including
losses and detection efficiency, the probability of a photon continuing after one step is typically around  50 $\%$  (without the EOM)  \cite{SchreiberScience}. Figure 7 (b) illustrates the projection of the 2D spatial lattice onto a 1D temporally encoded pulse
chain, for step one. Each step consists of a shift in both x-direction,
corresponding to a time difference of $\Delta t_{x}$, and y-direction corresponding to a time difference of $\Delta t_{y}$. Time delays are, in turn, adjusted by tuning the lengths of the SMFs. Time delay $\Delta t_{x}$ is adjusted by tuning the SMF length difference $|L_{2}-L_{1}|$, while $\Delta t_{y}$ is determined by the overall length $L_{1,2}$. Choosing $ \Delta t_{y} >> \Delta t_{x}$ it is possible to generate time bins characterized by distinctive arrival times. Figure 7 (c) depicts three different DTQW protocols that can be implemented using a similar experimental setup. Namely, (1)  Hadamard Quantum Walk (HQW), (2) Non-Commuting Rotations Quantum Walk (NCRQW), and (3) Split Step Quantum Walks (SSQW) (further details are in the text).\\

Note that the proposed experimental scheme only detects probability distributions encoded in arrival times for time bins in 2D DTQWs, but it does not retrieve phase information, which would be required to measure the Zak phase. Nevertheless, a  simple experiment to measure Zak phase differences for a closed path can be envisioned, based on the experimental scheme used to reconstruct holonomic phases, reported in \cite{HolonomicWhite}. Moreover, it is well known that the Zak phase is not a topological invariant itself, as the Berry connection is not gauge invariant under gauge transformations, in close analogy to the EM vector potential. Nevertheless, we note that the definition of the Bloch states via Eq. (2.4) does not specify the overall phase of $|u_{k} \rangle \rightarrow e^{i\chi(k)}|u_{k} \rangle $, so one can freely choose such phase. The singlevaluedness of $\chi(k)$ at the beginning and the end of the path imposes that the Zak phase for a given closed path is gauge invariant modulo $2 \pi$ \cite{Ozawa2018}.

\section{Discussion}

We have reported numerical results for calculation of the 2D Zak phase landscape in Discrete-Time Quantum Walk (DTQW) protocols, which can be readily  implemented in time-multiplexed photonic quantum walks \cite{SchreiberScience}. In particular, we investigated quantum walks which are driven by separable time-evolution operators of the form $U_{x} \otimes U_{y}$, which result in independent Bloch eigenvectors for each dimension ($x,y$). 
This warrants that the Berry curvature of the system ($F= \nabla \times A$) is always zero, where $A$ is the Berry connection whose integral over the first Brillouin zone yields the Zak phase. Nevertheless, more complex topological scenarios, with non-vanishing Berry curvature, can be explored by considering non-separable coin operations, via the implementation of controlled gates. Controlled gates condition the transformation of one coin state on the state of the other coin, thus introducing quantum correlations between the two dimensions. Because
of the induced quantum correlations, it is possible to obtain a non-trivial 2D evolution resulting in an inseparable final state, and a different acquired Zak phase for the system altogether, enabling to analyse the impact of quantum correlations and entanglement in topological structures, among other complex phenomena. Such exciting scenarios would enable to investigate the connection  between entanglement, Zak phase, Berry curvature, and Berry connection, and will be explored in upcoming works. Our results bear a close analogy to the Aharonov-Bohm effect, which essentially confirms that in a field-free ($F = 0$) multiply-connected region of space, the physical properties of the system depend on vector potentials ($A$), due to the fact that the Schroedringer equation is obtained from a canonical formalism, which cannot be expressed in terms of fields alone \cite{AharonovBohm1959}. 
In addition, our proposed protocols for exploration of 2D Zak phase landscape can be generalized to higher dimensions, by the introduction of orbital angular momentum, instead of polarization as the coin degree of freedom, and by adding extra temporal loops to encode the extra dimensions. To a large extent, this renders a fully unexplored avenue of research, enabling
quantum simulation applications with multiple walkers and nonlinear interactions, in high dimension. It may be possible
to study the effects of higher-dimensional graph percolations, localization effects, or the use the quantum  network
topologies in conjunction with single-photon or multi-photon 
states.

\begin{acknowledgements}
The Author is grateful to Alberto Grunbaum, Janos Asboth, and Osvaldo Santillan for many insightful discussions.   G P. acknowledges financial support from PICT2015-0710 StartUp grant and Raices Programme. 
\end{acknowledgements}

\end{document}